\documentclass[12pt]{article}
\usepackage[dvips]{graphicx}

\begin{document}
\title{Consequences of Partial Vector Meson Dominance for the
  Phenomenology of  Colored Technihadrons}
\author{ Alfonso R. Zerwekh
  \footnote{alfonsozerwekh@uach.cl} \\
 Instituto de F{\'{\i}}sica, Facultad de Ciencias \\
Universidad Austral de Chile \\ Casilla 567, Valdivia, Chile}
\date{}

\maketitle

\begin{abstract}
  In this paper we address the question: ``do the limits on technirho
  production at the Tevatron mean what we think they do ?'' These
  limits are based on calculations that rely on  
  Vector Meson Dominance (VMD). VMD was invented in order to describe the
  interaction of electrons with hadrons (the rho meson and pions). The
  method has been used also as a tool in the study of Technicolor
  phenomenology. Nevertheless there are evidences in the sense that,
  even in its original, 
  context VMD is not completely realized. In this work we investigate
  the consequences of a deviation from complete VMD for the
  phenomenology of colored technihadrons. We focus specially on the
  production of the color octet technirho and color triplet
  technipions. We found that a relative small direct coupling of the
  proto-technirho to quarks is enough to suppress or even eliminate the
  interaction among quarks and the physical technirho. On the other
  hand, it is possible to suppress the coupling of the physical technirho to
  technipions but in this case a large interaction among the
  technipions and the proto-gluon must be introduced. The
  consequences for the limits on the mass of the color octet technirho
  and colored triplets technipions are also investigated.     
   
\end{abstract}

\section{Introduction}
\label{sec:intro}

We do not know what mechanism has been chosen by Nature to break the
electroweak symmetry. There is a general agreement in the sense that the
Higgs sector of the Standard Model is incomplete and new Physics must
exist at the TeV scale in order to explain the origin of electroweak
symmetry breaking. An appealing possibility is that a new
strong interaction breaks the electroweak symmetry through the
formation of a condensate of new fermions which is not a singlet of
the electroweak symmetry group. This hypothetical new interaction is
generally called Technicolor\cite{technicolor}. In general,
Technicolor models have a rich 
spectrum of composite pseudoscalar and vector states. Some non-minimal
models predict the existence of colored technihadrons such as color
octet technirhos and color triplet technipions. They are the
focus of this work.

Because Technicolor is a strongly coupled theory, it is necessary to
use effective methods in order to study its phenomenology. In general,
the methods that have been proved useful in the description of the
properties and interactions of usual hadrons are also used in the
study of technihadrons. One of such methods is Vector Meson Dominance
(VMD)\cite{VMD}. A
related, but perhaps more elegant, method for describing such
interactions is the so called Hidden Symmetry. A realization of this
technique applied to a strong electroweak sector is known as the BESS
model\cite{dominici}.

VMD was invented in order to describe the interaction of electrons
with hadrons (the rho meson and pions). In this scheme the electron
interacts directly only with the (proto-)photon while the pion
interacts directly only with the (proto-)rho meson. The interaction
between electrons and pions is made possible by a mixing of the
(proto-)photon with the (proto-)rho. In a similar way, it is assumed
that the normal fermions interacts with technihadrons through a mixing
of the Standard Model gauge bosons with the vector technimesons.

Nevertheless, some deviations from complete VMD occur in
its original context\cite{Schechter:1986vs,myvmd}. For example, it has
been claimed that the decay of the usual (QCD) $\rho(770)$ into a pair
of leptons
is better described if a direct coupling of the leptons to the $\rho$
is introduced\cite{Schechter:1986vs}. On the other hand, in
\cite{myvmd} it is shown that the introduction of a direct coupling of
the photon to the charged pions, allows the abandon of the
universality condition in order to obtain a correct pion form
factor. Moreover, we know that Technicolor dynamics is not like
QCD. All these circumstances motivate us to ask about the meaning of
the limits on technirho production at the Tevatron when departure from
traditional VMD are consired.
 
In this work we investigate the consequences of 
deviations form  complete VMD (which we will call partial VMD) for the
production of a color octet technirho and a pair of color triplet
technipions.

\section{General Framework}
\label{sec:model}

The first studies of the color octet technirho phenomenology 
\cite{Lane:1991qh,Eichten:1994nc} were done using a version of VMD
(generalized  to color interaction) based on non-diagonal
propagators. In this work, we start with a version of VMD which is
equivalent to the previous one but is formulated in terms of
mass-mixing terms. 

As it was previously shown \cite{Zerwekh:2001uq}, the mixing of
the color octet technirho with the gluon can be described by the
Lagrangian:
 
\begin{equation}
  \label{eq:L1}
  {\cal L}=-\frac{1}{4}\tilde{G}_{\mu \nu}^a \tilde{G}^{a \mu \nu}
-\frac{1}{4}\tilde{\rho}_{\mu \nu}^a \tilde{\rho}^{a \mu \nu}
+\frac{1}{2}M_G^2\tilde{G}^2
+\frac{1}{2}M_{\tilde{\rho}}^2\tilde{\rho}^2
-\frac{\tilde{g}}{g'}M_{\tilde{\rho}}^2\tilde{G}\tilde{\rho}.
\end{equation}
where
\begin{eqnarray*}
   \tilde{G}_{\mu
  \nu}^a&=&\partial_{\mu}\tilde{G}^a_{\nu}-\partial_{\nu}\tilde{G}^a_{\mu}
-\tilde{g}f^{abc}\tilde{G}^b_{\mu}\tilde{G}^c_{\mu}\\
 \tilde{\rho}_{\mu
  \nu}^a&=&\partial_{\mu}\tilde{\rho}^a_{\nu}-\partial_{\nu}
\tilde{\rho}^a_{\mu} 
-g'f^{abc}\tilde{\rho}^b_{\mu}\tilde{\rho}^c_{\mu}.
\end{eqnarray*}

This Lagrangian is invariant under $SU(3)_C$ provided that the fields
transform as:

\begin{eqnarray}
  \label{eq:trans}
   \delta
   \tilde{G}^a&=&-f^{abc}\tilde{G}^b\Lambda^c-\frac{1}{\tilde{g}}
\partial\Lambda^a \nonumber \\
 \delta \tilde{\rho}^a&=&-f^{abc}\tilde{\rho}^b\Lambda^c-\frac{1}{g'}
\partial\Lambda^a.     
\end{eqnarray}

Notice that the fields $\tilde{G}$ and $\tilde{\rho}$ are not physical
fields because they are not mass eigenstates. We call them the proto-gluon and
the proto-technirho, respectively. The physical fields can be written
as :

\begin{eqnarray}
\label{eq:campfisicos}
  G_{\mu}^a&=&\tilde{G}_{\mu}^a\cos \alpha + \tilde{\rho}^a \sin
  \alpha \nonumber \\
\rho_{\mu}^a&=&-\tilde{G}_{\mu}^a\sin \alpha + \tilde{\rho}_{\mu}^a \cos
  \alpha
\end{eqnarray}
where $\alpha$ is given by 
\begin{equation}
  \sin\alpha = \frac{\tilde{g}}{\sqrt{{g'}^2+\tilde{g}^2}}.
\end{equation}
As usual, we estimate the value of $\alpha$ scaling up the mixing
between the photon and the usual rho meson obtained in normal VMD:

\begin{equation}
\label{eq:sina}
  \sin\alpha = \frac{g}{\sqrt{2\pi}}\frac{1}{\sqrt{2.97\times3/N_{TC}}}
\end{equation}
where $N_{TC}$ is the number of technicolors. As usual, we set
$N_{TC}=4$.

When the Lagrangian (\ref{eq:L1}) is written in terms of the physical
fields, we find that there is no coupling between a color octet
technirho and two gluons \cite{Zerwekh:2001uq}. Nevertheless,
it has been shown \cite{SekharChivukula:2001gv} that an operator with
dimension six 
exits that restores this coupling. This operator
may be written as:

\begin{equation}
  \label{eq:op6}
  {\cal
    L}=i\frac{c_2}{4\pi\Lambda^2_{TC}}f^{abc}\rho^{a\mu\nu}G^{b\gamma}_{\mu}
G^{c}_{\nu \gamma}
\end{equation}

Of course we do not know the value of the constant $c_2$ and hence an
important uncertainty exists about the contribution of this
operator. For simplicity, we work in
the  pessimistic limit assuming that the constant $c_2$ is small enough to
make this contribution negligible. In fact, at the Tevatron this term
is not important due to the low gluon luminosity at large partonic
center-of-mass energies \cite{Zerwekh:2004vy}. However, this term is
crucial for the study of the color octet technirho phenomenology at
the LHC. In this case, a reliable estimation of $c_2$ based on dynamical
calculations would be very valuable, but this analysis is beyond the
scope of this work.

\section{Color Octet Technirho Coupling to Quarks}
\label{sec:rhoqq}

Now we want to couple these fields to quarks. In normal VMD, this is
done assuming that the quarks only interacts with the
proto-gluon. Nevertheless, because both, the proto-gluon and the
proto-technirho, transform like gauge fields, it is possible to write
a covariant derivative with both fields:

\begin{equation}
\label{eq:Lag_q_gauge}
  {\cal L}=\bar{\psi}i\gamma^{\mu}\left[\partial_{\mu}
  -i(1-x)\tilde{g}\tilde{G}^a_{\mu}\frac{\lambda^a}{2}
  -i x g'\tilde{\rho}_{\mu}^a\frac{\lambda^a}{2} 
\right]\psi.   
\end{equation}
In this way, we introduce a direct coupling between the quarks and the
proto-technirho which strength is measured by the parameter $x$. This
kind of direct coupling can be produced by Extended Technicolor
although in this case it must be proportional to the quark mass and,
hence, small. Nevertheless because we do not know exactly all the
properties of the underlying (Extended) Technicolor theory, we can not
ignore the possibility that a more important coupling can be generated.        

When Lagrangian (\ref{eq:Lag_q_gauge}) is written in terms of physical
fields we obtain:

% \begin{equation}
%   {\cal L}=i\bar{\psi}\gamma^{\mu}\partial_{\mu}\psi +
%   g\bar{\psi}\gamma^{\mu}G^a_{\mu}\frac{\lambda^a}{2}\psi +
%   \frac{g}{\cos\alpha
%   \sin\alpha}(x-\sin^2\alpha)\bar{\psi}\gamma^{\mu}\rho^a_{\mu}
% \frac{\lambda^a}{2}\psi   
% \end{equation}

\begin{equation}
  {\cal L}=i\bar{\psi}\gamma^{\mu}\partial_{\mu}\psi +
  g\bar{\psi}\gamma^{\mu}G^a_{\mu}\frac{\lambda^a}{2}\psi +
  g\tan\alpha
  (\frac{x}{\sin^2\alpha}-1)\bar{\psi}\gamma^{\mu}\rho^a_{\mu}
\frac{\lambda^a}{2}\psi   
\end{equation}
where $g=\tilde{g}\cos\alpha= g' \sin\alpha$ is the usual QCD coupling
constant. Notice that the coupling of the gluon to quarks is the
usual one and it is independent of $x$. On the other hand, the
coupling of the physical color octet technirho not only depends on
$x$, it vanishes for $x=\sin^2\alpha$. That means that a relative
small deviation from normal VMD can produce the decoupling of the
physical technirho from quarks. Of course this effect has important
consequences for the present mass limits obtained at the Tevatron.

\section{Dijets Production at the Tevatron}
\label{sec:dijets}

In order to evaluate the effects, on the technirho phenomenology, of such
a modification of the technirho-quark interaction, we compute (with
the help of LanHEP \cite{LanHEP} and CalcHEP \cite{CalcHEP}) the cross
section for the resonant production of dijets at the Tevatron and we
compare it with the experimental limits obtained by the CDF collaboration
for a luminosity of $106$ pb$^{-1}$ (Run I)\cite{Abe:1997hm}. We use the CTEQ6L
\cite{cteq} parton distribution function, $\sqrt{s}=1800$ GeV  and we
implement the following kinematical cuts: 

\begin{equation}
  \label{eq:cutseta}
  |\eta|<2.0  \mbox{          for both jets}
\end{equation}
and 
\begin{equation}
  \label{eq:cutscost}
  |\cos\theta^*|=\left|\tanh\left(\frac{\eta_1-\eta_2}{2}\right)\right|<\frac{2}{3}        
\end{equation}
where $\eta$ is the jet pseudo-rapidity.

\begin{figure}[htbp]
  \centering
  \includegraphics{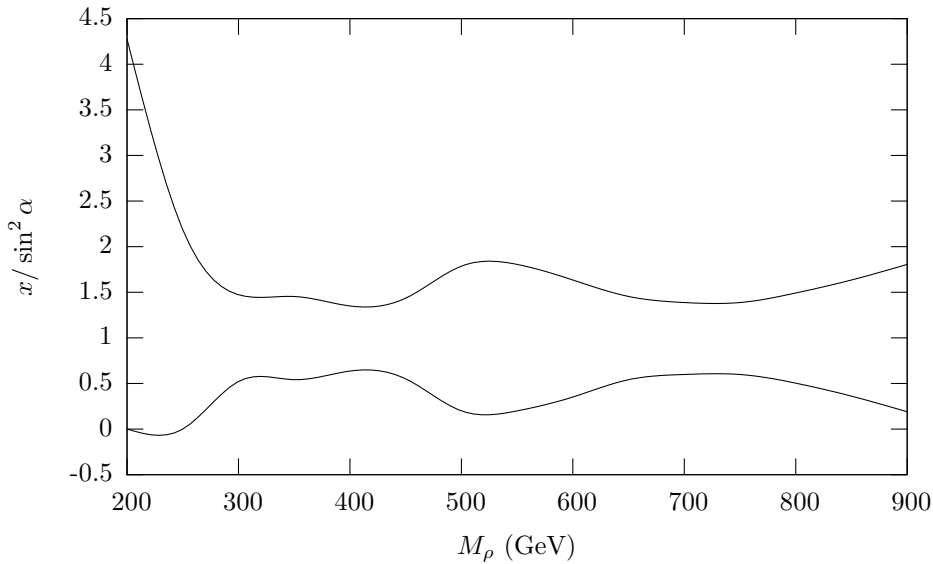}
  \caption{Limits on the $x$ parameter. The region between the lines
  is allowed}
  \label{fig:x}
\end{figure}

 As we have already said, we work in the limit where the
color octet technirho does not couple to two gluons, hence it is
produced only by quark fusion. For this reason, we do not interpret
the experimental results as limits on the color octet technirho mass,
but as limits on the $x$ parameter. The results are shown in figure
\ref{fig:x}. The curves represent the $95\%$ C.L. limits on $x/\sin^2
\alpha$ and the region between them is allowed.  

% Notice that a value of $x \approx 0.07$ is enough to render the
% color octet technirho invisible in its whole mass range.

%%%%%%%%%%%%%%%%%%%%% Answer to Question 1 %%%%%%%%%%%%%%%%%%%%%%%%
The color-octet technirho remains invisible in this channel if $x/\sin^2
\alpha \approx 1\pm 0.5$. For the value of $\sin\alpha$ given by
equation (\ref{eq:sina}), we obtain that the value of $x$ needed in
order to explain the non-observabilty of the color-octet technirho at
the Tevatron is $x\approx 0.11 \pm 0.05$. This value of $x$ seems too
high because it implies that ETC would produce unacceptable big masses
for the $u$ and $d$ quarks. Nevertheless, this numerical result
depends of the expression of $\sin\alpha$ shown in equation
(\ref{eq:sina}) that was scaled up from QCD. This procedure is highly
dangerous because we know that Technicolor's dynamics is not QCD-like.
Of coure, we can ask  whether, in a realistic model, the mixing angle
can be small enough in order to have an $x \approx \sin^2 \alpha$
compatible with the $u$ and $d$ quarks. To our knowledge, this is an
open question.

On the other hand, in the context of the top quark production in the so
called Extended Bess model \cite{EBESS} it has been considered a
direct coupling of a color-octet vector resonances (equivalent to our
color-octet technirho) to quarks of the same order of magnitude of the
lower bound of our $x$.

%%%%%%%%%%%%%%%%%%%%%%%%%%%%%%%%%%%%%%%%%%%%%%%%%%%%%%%%%%%%%%%%%%%

\begin{figure}[htbp]
  \centering
  \includegraphics{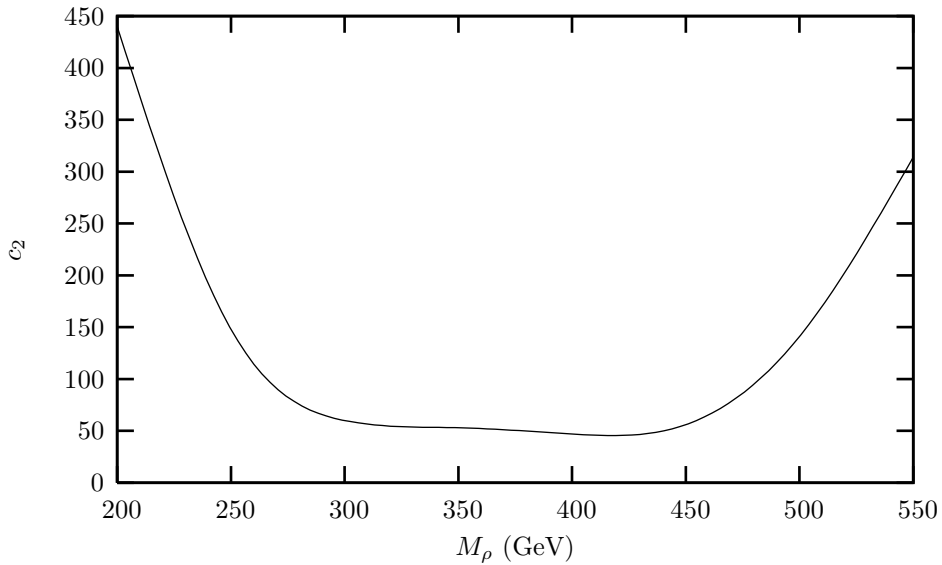}
  \caption{Limits on the $c_2$ parameter. The region below the curve
  is allowed}
  \label{fig:c_2}
\end{figure}

Let us turn again our attention to the color octet technirho
production by gluon fusion, governed by the dimension 6 operator
showed in equation (\ref{eq:op6}). We have already said that this
process is subdominant at the Tevatron due to the low gluon luminosity
at large partonic center-of-mass energies. Nevertheless, it necessary
to study its contribution when the coupling of the technirho to quarks
is suppressed. In this case we compute the production cross using
$\Lambda_{TC}=1 $ TeV and we compare our results with the experimental
limits on the production of a color octet technirho, in order to
obtain limits on the parameter $c_2$. The results are shown in figure
\ref{fig:c_2}. The region below the curve
  is allowed

Notice that even for large values of $c_2$ the color octet technirho
remains invisible in the dijet spectrum. We recall that, on the ground
of dimensional analysis, the natural value of $c_2$ is of ordrer $O(1)$

\section{Technipions Pairs Production}
\label{sec:lq}

Another interesting process is the production of colored
technipions. In this section we focus on the production of the color triplet
ones. In the One Family Technicolor Model, there are four color
triplet technipions which we denote generically by $\pi_3$. We assume
that all of them have the same mass. In an exact analogy with the
previous case, the Lagrangian may be written in terms of our extended
covariant derivative: 

\begin{eqnarray}
  \label{eq:LQlagrangian}
  {\cal L}= D_{\mu}\pi_3 
   \left(D^{\mu} \right)^{\dagger}\bar{\pi}_3-M_{\pi}^2\pi_3 \bar{\pi}_3 
\end{eqnarray}
where
\begin{equation}
  \label{eq:Dlq}
  D_{\mu}=\partial_{\mu}-(1-y)i g'
  \frac{\lambda^a}{2}\tilde{\rho}^a_{\mu}-y i \tilde{g}
  \frac{\lambda^a}{2} \tilde{G}^a_{\mu}
\end{equation}
In this case, the parameter $y$  measures the strength of the direct
coupling between the proto-gluon  and the color triplet technipion. The
relevant Feynman
rules for the technipion pair production, written in terms of the
physical fields, are shown in table \ref{tab:FR} . 

\begin{table}[htbp]
  \centering
 \caption{Feynman Rules relevant for the technipion pair production. }
\begin{tabular}{l l} 
\hline
\hline
Fields in the vertex & Variational derivative of Lagrangian
\\ \hline
${G}_{\mu p }$ \phantom{-} ${\bar{\pi}}_{3q }$ \phantom{-} ${\pi}_{3r
}$ \phantom{-}  & $ g \frac{\displaystyle \lambda_{q
    r}^p}{\displaystyle 2} \big(p_3^\mu -p_2^\mu
\big)$\\[2mm] 
${G}_{\mu p }$ \phantom{-} ${G}_{\nu q }$ \phantom{-} ${\bar{\pi}}_{3r }$
\phantom{-} ${\pi}_{3s }$ \phantom{-}  & $ g{}^2 g^{\mu \nu}
\big(\frac{\displaystyle \lambda_{r t}^p}{\displaystyle 2}
\frac{\displaystyle \lambda_{t s}^q}{\displaystyle 2} +
\frac{\displaystyle \lambda_{r t}^q }{2} \frac{\displaystyle
  \lambda_{t s}^p}{\displaystyle 2}
\big)$\\[2mm]  
${\rho}_{\mu q }$ \phantom{-}${\bar{\pi}}_{3p }$  \phantom{-}
${\pi}_{3r }$ \phantom{-}  & $\frac{\displaystyle g}{\displaystyle  \cos\alpha
  \sin \alpha}(\cos^2\alpha-y)\frac{\displaystyle \lambda_{p
    r}^q}{\displaystyle 2} \big( p_3^\mu -p_2^\mu 
\big)$\\[2mm]   
\hline
\hline
  \end{tabular}
  \label{tab:FR}
\end{table}

Notice that the couplings of the physical gluon to technpions are
independent of $y$ and are exactly the same ones we  would obtain in scalar
QCD. On the other hand, the coupling of the physical technirho depends
on $y$ but it is suppressed only if $y\approx \cos^2 \alpha$. That is,
a large proto-gluon to technipions coupling is needed, what, in
principle, seems unnatural.

\begin{figure}[htbp]
  \centering
  \includegraphics{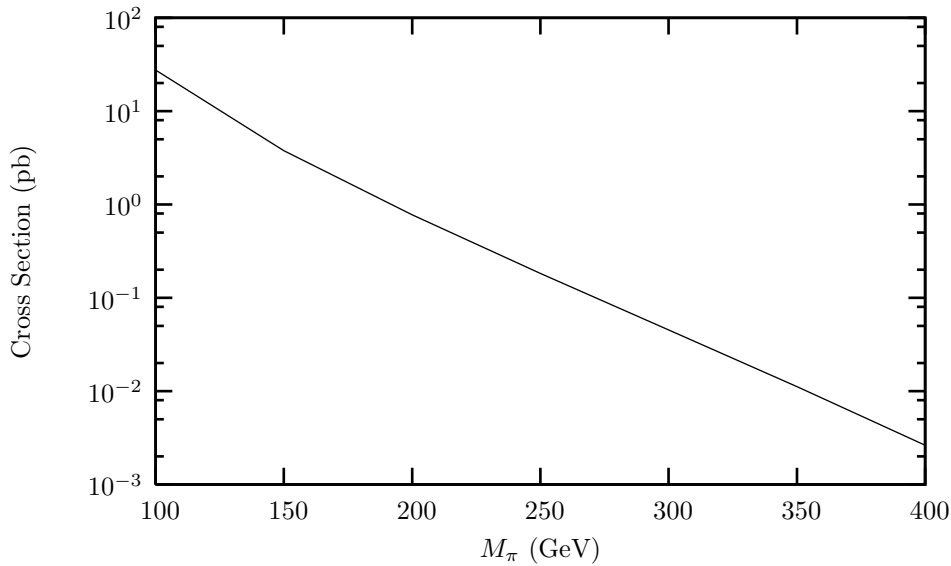}
  \caption{Non-resonant color triplet technipion pair production cross
  section as a function of the technipion mass for $\sqrt{s}=1800$ GeV
  and $y=0$. }
  \label{fig:lq}
\end{figure}

Of course, the technirho production still can be suppressed by the
mechanism discussed in the previous section, and then the technipion
pair production would become non-resonant. Under this conditions, we
compute the technipion pair production cross section for
$\sqrt{s}=1800$ GeV as a function of the technipion mass. The result
is shown in figure \ref{fig:lq}. In general, the present limit for the
production cross section of a pair of color triplet technipions is
$\sigma(p\bar{p} \rightarrow \pi_{3}\bar{\pi}_{3}) < 0.5$ pb
\cite{Affolder:2000ny} and our results satisfy this limit for $M_\pi \geq
200$ GeV. Of course, the same result (i.e. figure \ref{fig:lq}) would
be obtained, without any assumpyion on the parameter $x$, if
$M_{\rho}<2M_\pi$ or if the $\rho$ is too heavy to be produced.

\section{Searching the Color-Octet Technirho at the LHC}

\begin{figure}[tb]
  \centering
  \includegraphics{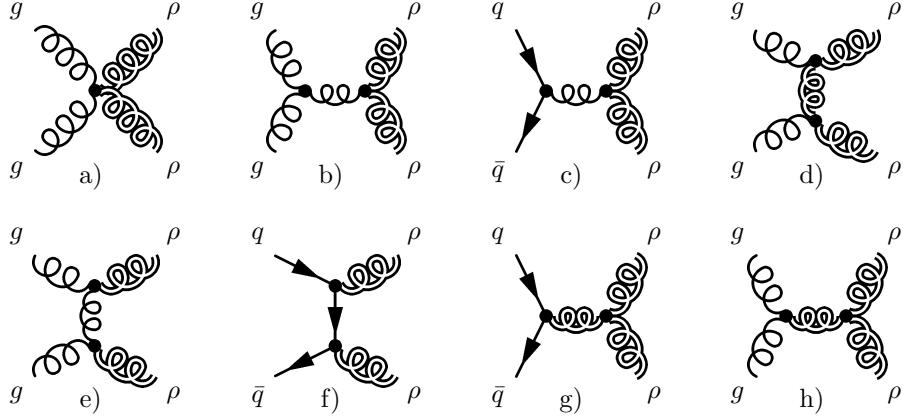}
  \caption{Feynman diagrams for the production of a pair of
    color octet technirhos at the LHC. }
  \label{fig:lhc_diag}
\end{figure}

Let us turn, for a moment, our attention to the LHC.
As we have seen in the previous sections, the production of a color
octet technirho suffers of significants theoretical
uncertainties. But the LHC will offer a center of mass energy and a
luminosity big enough to consider other channels like the production
of a pair of color octet technirhos. The Feynman diagrams for this
process are shown in figure \ref{fig:lhc_diag}. The last four diagrams
(from e to h) depend on the mixing angle, $c_2$ and the $x$ parameter,
in contrast to the first four diagrams (from a to d) which are model
independent. In fact, the only couplings that participate in  the
first four diagrams  of figure \ref{fig:lhc_diag} are (except for
usual QCD) $G\rho\rho$ and $GG\rho\rho$ and they are dictated by gauge
invariance. The Feynamn rules for these couplings can be obtained from
the Lagrangian (\ref{eq:L1}) and are presented in table \ref{tab:GR}.

\begin{table}[htb]
  \centering
  \caption{Feynman rules relevants for the technirho pair production}
\begin{tabular}{ll}
\hline
\hline
Fields in the vertex & Variational derivative of Lagrangian by fields \\ \hline
${G}_{\mu p }$ \phantom{-} ${R}_{\nu q }$ \phantom{-} ${R}_{\rho r }$ \phantom{-}  &
	$gf_{p q r} \big(p_1^\rho g^{\mu \nu} -p_3^\mu
        g^{\nu \rho} +p_3^\nu g^{\mu \rho}$\\
  & $-p_1^\nu g^{\mu \rho}+p_2^\mu g^{\nu \rho} -p_2^\rho g^{\mu \nu} \big)$\\[2mm] 
${G}_{\mu p }$ \phantom{-} ${G}_{\nu q }$ \phantom{-} ${R}_{\rho r }$
\phantom{-} ${R}_{\sigma s }$ \phantom{-}  &
	$g{}^2 \big(g^{\mu \rho} g^{\nu \sigma} f_{p q t} f_{r s t}
        +g^{\mu \nu} g^{\rho \sigma} f_{p r t} f_{q s t} -g^{\mu
          \sigma} g^{\nu \rho} f_{p r t} f_{q s t} $ \\[2mm]
  & $-g^{\mu \sigma} g^{\nu \rho} f_{p q t} f_{r s t} +g^{\mu \nu}
  g^{\rho \sigma} f_{p s t} f_{q r t} -g^{\mu \rho} g^{\nu \sigma}
  f_{p s t} f_{q r t} \big)$\\ 
\hline
\hline
\end{tabular}
 \label{tab:GR}
\end{table}

\begin{figure}[tb]
  \centering
  \includegraphics{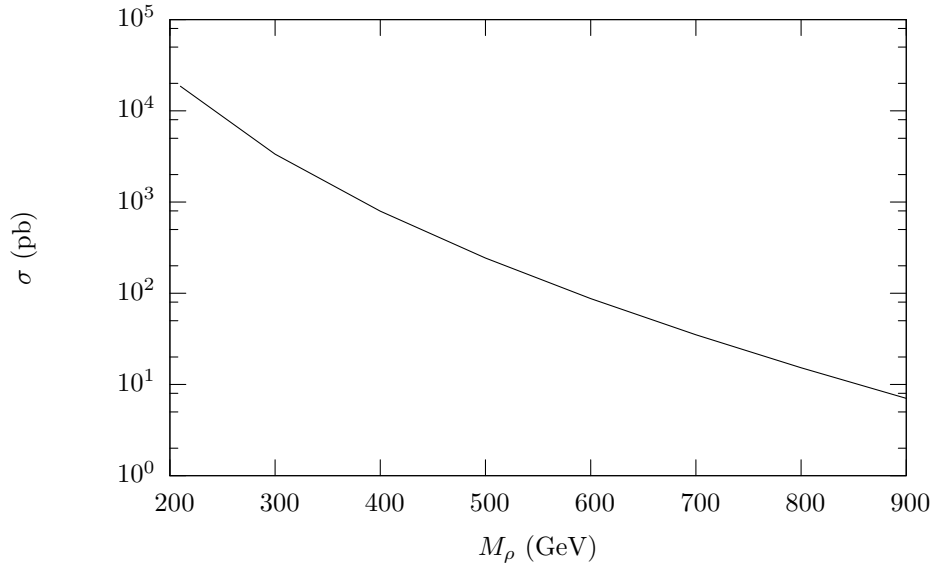}
  \caption{Pair production of color octet technirhos at the LHC
    through gluon fusion. Remark that we use $c_2=0$}
  \label{fig:lhc}
\end{figure}

We estimate the pair production of color octet technirhos using only
the diagrams \rm{a}, \rm{b} and \rm{d} of fig. \ref{fig:lhc_diag} because
we expect that the dominant contribution to this process, at the LHC,
comes from gluon fusion. The result
is shown in figure \ref{fig:lhc}. It is expected that LHC will have a
luminosity of ${\cal L}=10^{4}$ pb$^{-1}$/yr so the color octet
technirhos must be abundantly produced (between $10^8$ and $10^4$
events per year in the whole possible mass range) at the LHC. Of course, a more
detailed analysis taking into account realistic detector effects and background
must be done, but such a work is beyond the scope of this paper.  

\section{Conclusion}
\label{sec:conclusion}
We have written down the more general Lagrangian (using operators with
dimension up to four) that describes the
interaction of the the color octet technirho with quarks, gluons and
colored technipions. Our Lagrangian deviate from the usual
implementation of Vector Meson Dominance because it includes a direct
coupling of the technirho to quarks. We found that the coupling of the
physical technirho to quarks is modified in such a way that it can be
significatively suppressed with respect to the usual expectations of
normal VMD. In fact, we compared our predictions to experimental data
from the CDF Collaboration and we found that a relative small direct
coupling can render the color octet technirho invisible. This effect
has consequences also for the pair production of colored technipions,
which becomes non-resonant. Nevertheless, the data obtained at the
Tevatron exclude the existence of color triplet technipions for $M_\pi
< 200$ GeV. Finally, we propose to pay attention to the pair
production of color octet technirhos as a viable discovery channel at
the LHC.
  
% The suppresion of the coupling of the technirho to quarks happens when
% $x \approx sin^2\alpha$. When we use the value of $\alpha$ scaled up from
% usual hadron phenomenology, we find $x \approx 0.1$. This value may
% seem to high to be produced by Extended Technicolor while keeping a
% low mass for the $u$ and $d$ quarks.

\section*{Acknowledgments}

The author received support from Universidad Austral de Chile (DID
grant S-2006-28).

\end{document}